\documentclass[letterpaper, 10 pt, conference]{ieeeconf}  

\IEEEoverridecommandlockouts                              

\usepackage{graphics} 
\usepackage{epsfig} 
\usepackage{mathptmx} 
\usepackage{times} 
\usepackage{amsmath} 
\usepackage{amssymb}  
\usepackage{CJKutf8}
\title{\LARGE \bf
IR Design for Application-Specific Natural Language: 、\\A Case Study on Traffic Data
}

\author{Wei Hu$^{1}$, Xuhong Wang$^{1}$, Ding Wang$^{1}$, Shengyue Yao$^{1}$, Zuqiu Mao$^{2}$, Li Li$^{3}$, \textit{Fellow, IEEE},
\\Fei-Yue Wang$^{4}$, \textit{Fellow, IEEE}, and Yilun Lin$^{1,*}$, \textit{Member, IEEE}
\thanks{*This work is supported by the Shanghai Artificial Intelligence Laboratory}
\thanks{*Corresponding author: Yilun Lin (linyilun@pjlab.org.cn)}
\thanks{$^{1}$Wei Hu (huwei@pjlab.org,cn), Xuhong Wang (wangxuhong@pjlab.org.cn), Ding Wang (wangding@pjlab.org.cn), Shengyue Yao (yaoshengyue@pjlab.org.cn), Yilun Lin (linyilun@pjlab.org.cn) are with Urban Computing Lab, Shanghai AI Laboratory, Shanghai, China.}%
\thanks{$^{2}$Zuqiu Mao (Maozuqiu@51world.com.cn) is with 51 WORLD, Shanghai, China.}%
\thanks{$^{3}$Li Li (li-li@tsinghua.edu.cn) is with the Department of Automation, Tsinghua University, Beijing, China.}
\thanks{$^{4}$Fei-Yue Wang (feiyue.wang@ia.ac.cn) is with the Institute of Automation, Chinese Academy of Sciences, Beijing, China, and the Macau Institute of Systems Engineering, Macau University of Science and Technology, Macau, China.}
}

\begin{document}
\begin{CJK}{UTF8}{gbsn}

\maketitle
\thispagestyle{empty}
\pagestyle{empty}

\begin{abstract}

In the realm of software applications in the transportation industry, Domain-Specific Languages (DSLs) have enjoyed widespread adoption due to their ease of use and various other benefits. With the ceaseless progress in computer performance and the rapid development of large-scale models, the possibility of programming using natural language in specified applications - referred to as Application-Specific Natural Language (ASNL) - has emerged. ASNL exhibits greater flexibility and freedom, which, in turn, leads to an increase in computational complexity for parsing and a decrease in processing performance. To tackle this issue, our paper advances a design for an intermediate representation (IR) that caters to ASNL and can uniformly process transportation data into graph data format, improving data processing performance. Experimental comparisons reveal that in standard data query operations, our proposed IR design can achieve a speed improvement of over forty times compared to direct usage of standard XML format data.

\end{abstract}

%
\IEEEpeerreviewmaketitle

\section{INTRODUCTION}

DSL (Domain-Specific Language) is a programming language specifically designed to provide a solution to problems within a particular application domain. Compared to general-purpose programming languages, DSLs are more direct, easy to understand, write, and maintain \cite{mernik2005and}. The importance of DSL lies in providing a more effective way to deal with problems within specific domains. Domain-specific languages allow the expression of domain-specific concepts to be more natural, and the code is more readable and maintainable. This is because DSLs are designed specifically to address domain-specific issues and are intimately related to these problems.

Considering the different application scenarios and purposes, the characteristics of DSLs used by various tools vary. From a usability and processing performance perspective, these DSLs tend to offer better language usability when they have higher levels of abstraction and syntax flexibility. However, this leads to increased processing complexity and lower computer performance. Current advancements in computer performance have surpassed the limitations of human abilities, which have not significantly advanced due to physiological barriers. JIT (Just-In-Time) and IR (Intermediate Representation) can help mitigate DSL performance issues \cite{click1995simple,leissa2015graph,guo2019towards}. Recently, with the development of Large Language Models (LLMs) such as ChatGPT, programming in natural languages has become possible. This Application-Specific Natural Language (ASNL) is in high demand because domain experts typically lack programming experience and often struggle to utilize DSL or other complex programming languages in certain applications. Furthermore, some transportation-specific applications require specialized data structures that can be designed to support natural language-based semantics, which provide more flexibility and freedom, but also cause the performance issue of computer processing of language to become more prominent.

\begin{figure}[htb]
\centering
\includegraphics[width=0.8\linewidth]{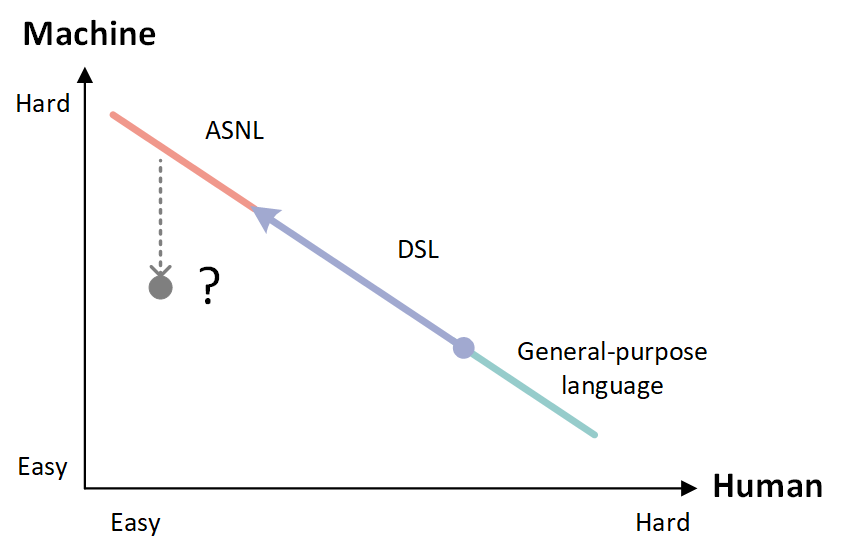}
\caption{The development trend of DSLs}
\label{fig_dsl2}
\end{figure}


DSLs are frequently utilized within transportation software to facilitate the work of professionals in this field. These DSLs specifically designed for processing traffic problems often focus on vehicle motion, route computation, traffic condition simulation, trajectory prediction, and navigation, and offer specific syntax and functionality tailored to these problems. Common transportation applications, such as SUMO \cite{krajzewicz2012recent}, Aimsun \cite{barcelo2005dynamic}, VISSIM \cite{fellendorf2010microscopic}, Paramics \cite{cameron1996paramics}, TRANSIMS \cite{smith1995transims}, Anylogic \cite{merkuryeva2010vehicle}, ROS \cite{yu2021mobile}, Apollo \cite{xu2019automated}, AutoWare \cite{kato2018autoware}, Simulink \cite{taylor2004macroscopic}, and Veins \cite{fernandes2012vns}, all have their own DSLs to facilitate the use for professionals outside of the computer field. Moreover, DSL in the field of transportation also has the tendency to develop further to the more free and flexible ASNL, such as openSenario proposed by ASAM. But these DSLs don't pay much attention to performance issues. Applications of DSL or ASNL in the transportation field have their own domain specific characteristics, the most significant of which is the involvement of extensive data operations, such as searching and modifying vast transportation facilities, road networks, and traffic flows. Improving the data processing capability is one of the key factors in enhancing the performance of ASNL in the transportation field.

There are two research methods in the field of autonomous driving simulation, known as worldsim and logsim \cite{creutz2021simulation,wang2023interpretable, feng2022application}. We can extend this concept to the application in the entire transportation field, which can also be categorized into two types of perspectives, namely worldview and logview. The former mainly focuses on the analysis of transportation log data, which is typically used for road condition analysis, traffic control, congestion relief, and other related applications. The latter, on the other hand, takes a macroscopic view with respect to the urban traffic situation from the perspective of geographic information systems (GIS), which can be used for traffic planning, road network design, heat map generation for traffic flow, and other related purposes.The DSLs for these two perspectives have completely different data structures. 


However, there is an urgent need to unify the data from these two perspectives, especially in the context of large-scale transportation models and comprehensive traffic data applications. Nevertheless, as mentioned earlier, some attempts in this area still lack significant attention to performance-related concerns.

This paper aims to resolve performance issues related to data processing in ASNLs by utilizing IR techniques. An IR is an abstract, machine-independent, low-level language representation that converts high-level languages into machine-executable code, while also being used for program optimization and code generation. We discuss the performance differences in large-scale transportation data queries using two different types of IRs: XML and Graph.  XML represents sequential form, while Graph represents structured form within the context of ASNL.  Our study demonstrates that structured, graph-based IRs outperform sequential ones in terms of information queries.  Graph-based IRs perform better in multi-party queries, synchronization, and node sharing as their structured form enables them to effectively address these challenges.
\begin{enumerate}
    \item \textbf{Referential Invariance:} Ensuring that data structures possess similar structure across different abstraction levels. Due to sizable gaps between natural language and DSL particularly in domains like transportation, semantic parsing and abstraction is necessary to ensure consistency in similar structural representation across different levels for compilation and optimization processes.
    \item \textbf{Performance Optimization:} Designing appropriate IR can effectively improve the processing performance of complex data in the ASNL. The IR can be designed to be independent of data sources and targets for facilitating cross-domain and cross-scale applications. The use of suitable abstract data structures is recommended to facilitate the adoption of parallelization and other optimization techniques. Optimization of data access techniques and caching strategies can help to reduce data access delays. Additionally, the analysis of data dependencies can be utilized to optimize data processing execution order and parallelization levels.
\end{enumerate}

In this paper we proposes a graph-based IR for ASNL. This approach enables the pre-processing of the various data structures utilized in different ASNL across diverse contexts and tools, into an IR using unified graph representation form, thereby effectively reducing the level of computational power necessary for subsequent information processing and exchange. In the minimal experimental case presented in this paper, a common vehicle location query task within the transportation domain was conducted. Compared to the commonly utilized XML data format, the speed of the Graph-based IR was increased by over 40 times.

\section{RELATED WORKS}
\subsection{Data format}
As previously mentioned, transportation applications can be broadly classified as either worldview or logview, which respectively focus on macro environmental factors and individual data records. Data formats in logview comprising primarily of raw or processed data in standard formats such as CSV, XML, and JSON obtained from various data sources, while worldview data formats pertain mainly to macro environmental factors such as urban infrastructure, road networks, and planning and use formats such as GIS, CAD, and specific XML formats. Logview emphasizes diverse data forms, while worldview necessitates standardized formatting standards to enhance efficient data interaction and analysis from disparate sources.

OpenStreetMap (OSM) is a prime example of a worldview perspective in the transportation domain. OSM is an open source map dataset that helps people better understand and use a city's transportation system by collecting a variety of open source data sources, such as government information, satellite imagery and user input, and then modeling and coding information about roads, buildings and more in an integrated map database \cite{wang2018data}. OpenStreetMap contains complete geographic information and utilizes the XML-based OSM format \cite{borkowska2022analysis}. OSM's XML structure is organized hierarchically, enabling scalability and readability. However, due to its design for enhanced readability rather than processing speed, parsing, processing, and editing data may require more time and computational resources when files are large, in order to alleviate this disadvantage, OSM simplifies some geometric information resulting in a certain loss of precision \cite{liebig2017dynamic}. OSM also lacks a data structure with direct support for searching and spatial queries, impeding efficiency when analyzing a large amount of tags and relationships.

DATEX II is another prominent example of a worldview perspective in transportation. Specifically, it is an international standard for exchanging road traffic data. It provides a generic data structure and information model for describing road networks, events, and journey information \cite{figueiras2017big}. However, DATEX-II may have slow processing speeds for large data sets due to its XML format, and it needs to be extended to suit different application requirements, which may result in gaps in managing different types of traffic information processes.

SUMO is a typical representative of the logview perspective in the transportation domain. Although it mainly centers on individual data such as vehicle positions, speeds, and routes, it also incorporates certain worldview elements in its modeling processes. Therefore, SUMO can be considered as a combination of logview and worldview perspectives in transportation research. SUMO employs CSV and XML formatted data to model and simulate traffic scenarios \cite{szalai2020mixed}.  However, the presence of numerous data element tags and a nested hierarchical structure can lead to larger file sizes and lower processing efficiency, making it more complex for editing and exchanging data between different software systems \cite{codeca2017towards}.

There have been some attempts to unify the different data formats in worldview and logview applications for a unified data format. The Association for Standardisation of Automation and Measuring Systems (ASAM) is a organization that aims to standardize the automotive industry by developing open data formats for automotive testing and simulation, and by utilizing two recommended formats OpenDRIVE and OpenSCENARIO, data from other applications can be unified. OpenDRIVE is used primarily for testing and simulation of advanced driver-assistance systems (ADAS) and automatic driving (AV) systems, offering features such as road geometry, lane topology, and time-variant road data \cite{chiang2022automated}. However, OpenDRIVE's complexity and potential data loss issues may require more training and skills for novices. In contrast, OpenSCENARIO is used to describe scenes for vehicle behavior and road environments \cite{chen2022generating}, making it well-suited for validating the function and performance of automotive automation systems. OpenSCENARIO 1.x is based on XML and shares features with OpenDRIVE, while OpenSCENARIO 2.x is a more specialized DSL with greater expressive power, enabling consistent scenario descriptions across different levels \cite{magnus2022virtual}. Nonetheless, OpenSCENARIO 2.x is still a new and not yet fully compatible DSL, requiring technical expertise and experience for effective scenario descriptions, and its road information still needs to refer to openDrive based on XML, for complex file processing performance has a short board.

\subsection{Performance improving}

As mentioned above, DSLs often employ data structures that exhibit suboptimal processing performance. One approach to rectifying this problem is to utilize JIT techniques \cite{lange2016devito}. In contrast with traditional compilers that generate binary files during compilation, JIT compilers dynamically compile the source code into machine code at runtime, allowing for code optimization during execution. The key to optimizing data structure performance using JIT lies in the implementation of a warm-up period that serves to minimize repeated interpretation of code, thus facilitating performance enhancement. During execution, the JIT compiler dynamically compiles the code and optimizes commonly used data structures based on the runtime environment and data conditions, converting them into persistent forms that mitigate the need for repeated interpretation and associated time consumption \cite{de2015just}.

IR is another possible technique for improving performance. Different data formats may require different techniques and algorithms for optimization. Using intermediate representation techniques, these algorithms and techniques can be easily migrated between different environments, thus improving the processing efficiency and performance of data formats. IR technologies have been widely applied in various fields. For instance, MLIR can be employed to improve the runtime efficiency of machine learning code \cite{bik2022compiler}. Developed by Google, MLIR serves as a bridge for high-performance computing and can compile different DSLs together, thereby eliminating the boundaries between these DSLs and enhancing the code efficiency. Furthermore, MLIR is capable of automatically optimizing code, allowing it to identify and optimize bottlenecks within the code, thus further improving its runtime efficiency. LLVM is a more general code performance optimization tool framework, and its code optimization is implemented through LLVM IR \cite{shajii2023codon}. LLVM IR is a low-level abstract syntax tree, in which the compiler decomposes the source code into low-level structures such as basic blocks and expressions so that code optimization and subsequent code generation can be performed during transmission. LLVM also supports JIT compilation technology to further improve code performance.

The performance issue also has been investigated from the perspective of data storage models, and the CD-DB storage model has been proposed to enhance the read and write performance of relevant data in collaborative driving applications \cite{yu2022cd}. CD-DB devised a data structure comprising of a skip list and a linked list for storing data of active vehicles, and a two-dimensional linked list for storing data of inactive vehicles, and the results of experiments showed that this approach can effectively improve the read and write performance of data in collaborative driving applications.

\section{METHODS}

\subsection{Domain Characteristics}

In light of the transition from DSL to more flexible and user-friendly ASNL, as discussed previously, there is a need to enhance the efficiency of program execution. To address this need, this article proposes the use of IR technology. High-performance and generic IR formats are generated and stored based on the data structure of various traffic simulation or autonomous driving software. This generic IR format can be utilized to convert or standardize different application data structures. On the one hand, the adoption of this generic IR can simplify the conversion and interaction of traffic data across different scales, domains, and sources in full scenarios. On the other hand, under conditions of frequent and extensive traffic data reading, writing, and modification, a high-performance IR format can effectively enhance computing efficiency and improve the real-time performance of software operation.

Designing a suitable IR demands careful consideration of the domain-specific characteristics of traffic applications. Traffic applications must process massive amounts of traffic infrastructure and scenario data, with the added challenge of dynamic traffic flows resulting in frequent lookup and modification operations. As previously noted, most current traffic data structures are based on XML, characterized by complex nested levels that are not user-friendly for frequent information analysis and processing. Furthermore, unlike the traditional approach of establishing separate models and simulations for specific applications and scales, the current trend of large-scale traffic modeling requires increasing demands for cross-scale and cross-domain data and interaction for all traffic scenarios. Standardized data structures often require mutual conversion between different standards, significantly reducing efficiency.

\subsection{Graph-based IR}

Based on these domain characteristics, this paper focuses on two key issues regarding the IR design in ASNL: referential invariance and performance optimization. Referential invariance refers to the ability of the data structure to maintain a consistent structure across different levels of abstraction. This is because there may be significant gaps between natural language and DSL, such as in the domain of transportation, where semantic parsing and abstraction may be necessary to ensure consistent representation of similar structures across different levels, thereby facilitating subsequent compilation and optimization processes. Performance optimization is another critical issue. In ASNL, we face the challenge of transforming natural language representations of applications into efficient machine code. To address this issue, IR design is required to support efficient compilation performance and optimization techniques.

The utilization of Graph-based IR can help address both referential invariance and performance optimization issues. With regard to referential invariance, graph representation can resolve correspondences between different-level data structures in DSL by representing natural language descriptions of traffic data as graphs. This ensures consistent representation of similar structures at different levels, thanks to the uniform structural form and semantic representation of graph representation. Additionally, graph representation's features can facilitate semantic inference and query operations, allowing traffic data processing and analysis. 

Performance optimization can be advantageous by leveraging graph representation. A unified graph data structure representation facilitates more effective data flow analysis and optimization. The key technologies in this process include data flow analysis, algorithm optimization, parallel computing, and cache optimization. Data flow analysis optimizes code execution sequence and parallelization level by analyzing the dependency relationships between data. Algorithm optimization focuses on crafting efficient compilation optimization algorithms for typical algorithms and operation patterns within ASNL to reduce computational and storage costs. Parallel computing is utilized to devise and implement parallel compilation and execution strategies, taking full advantage of the parallel computing capabilities of modern hardware to enhance application performance and responsiveness. Cache optimization optimizes data access techniques and cache strategies to reduce data access latency and increase program execution efficiency.

In practical application, the Graph data structure can not only describe basic information about road networks and roads, but can also include more attribute information, such as vehicle speed, traffic flow, road congestion, traffic signals, etc., while also having good scalability and portability \cite{czerepicki2016application}. The Graph data structure makes it easy to perform calculations such as route planning, traffic state prediction, and travel time optimization \cite{miler2014shortest}. Graph has strong capabilities for representing relationships between data and high computational efficiency, which is particularly important for the analysis of traffic data, where relationships play a crucial role. Improving processing performance in traffic simulation and autonomous driving applications is important \cite{chen2020graph}. At the same time, the Graph data structure maintains the characteristic of ASNL being simple and easy to understand for humans. Compared with complex nested XML data structures and CSV data structures, Graph structures are more easily recognized and understood by humans. 

\begin{figure}[htbp]
\centering
\includegraphics[width=0.8\linewidth]{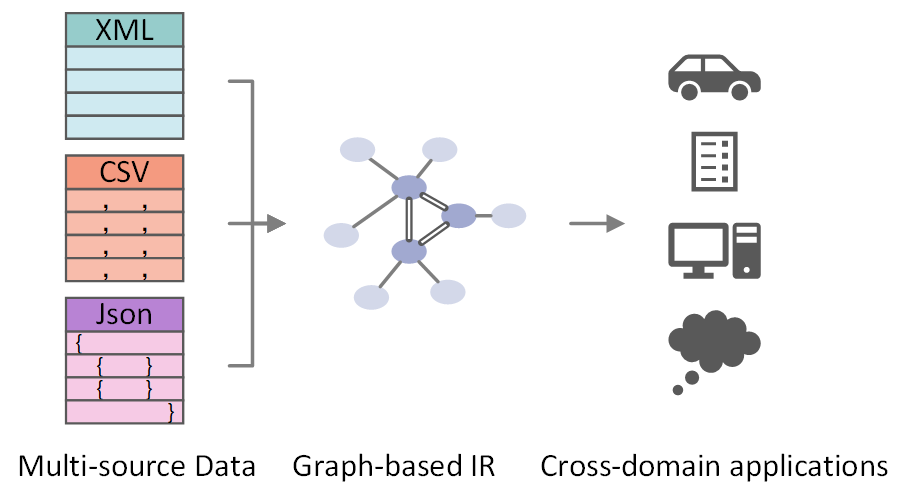}
\caption{Graph-based IR for multi-source data and cross-domain applications}
\label{fig_IR4ASNL}
\end{figure}

\begin{figure*}[htbp]
\centering
\includegraphics[width=0.65\linewidth]{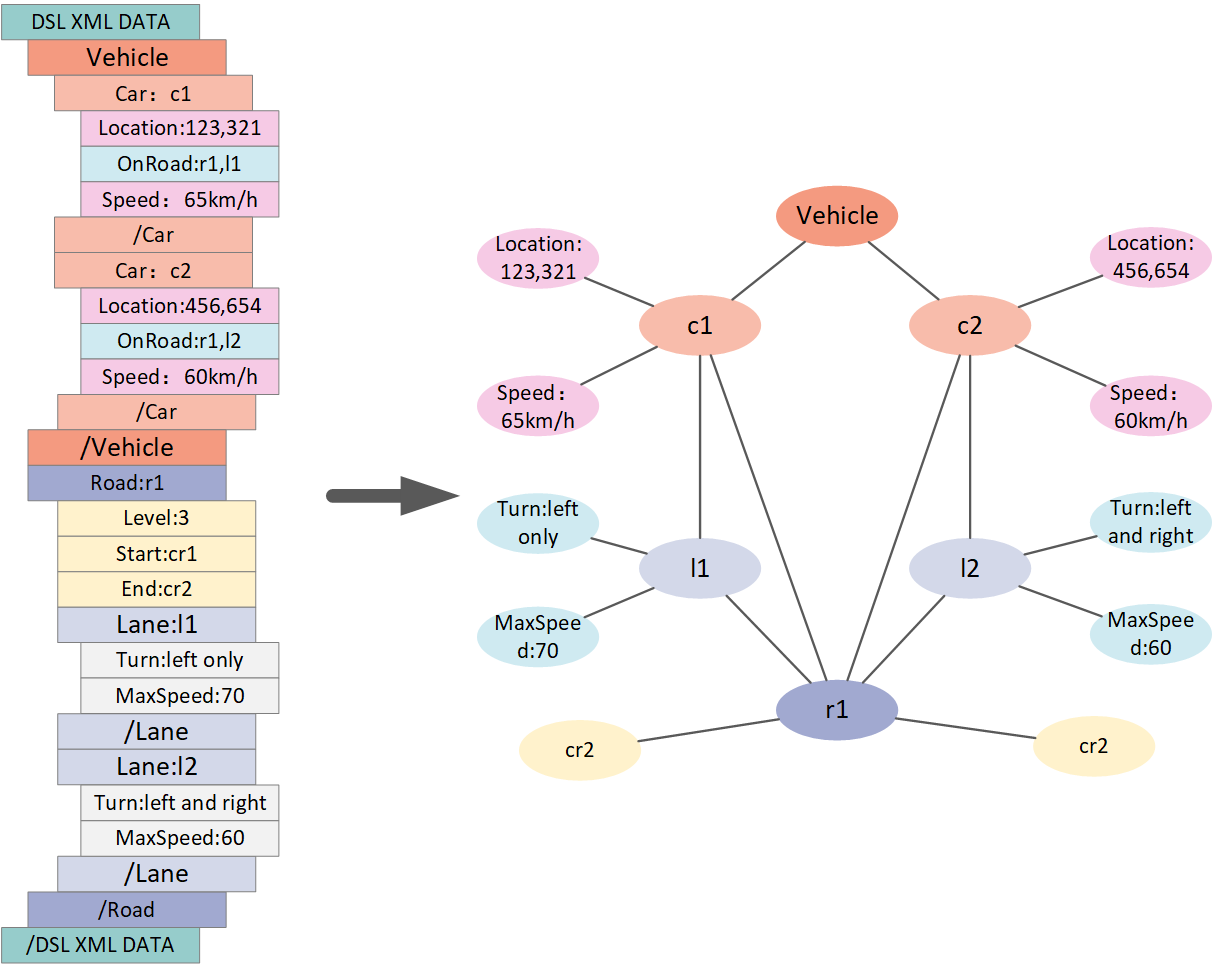}
\caption{The XML data format is converted to a unified Graph-based IR}
\label{fig_xml2dsl}
\end{figure*}

The method proposed in this paper uses Graph as an efficient and universal intermediate representation for ASNL data structures in the traffic domain. To enhance performance for frequently manipulated data, the conversion of text-based XML, CSV and Json data formats to a generic Graph-based IR can be achieved, as shown in Fig. \ref{fig_IR4ASNL}. Subsequently, the unified Graph-based IR is stored in a Graph database to improve the performance in subsequent application data exchange and processing. It also maintains human interpretability. As an example, Fig. \ref{fig_xml2dsl} illustrates the conversion of XML representation into a Graph-based IR. The left side of the diagram shows traffic scene data represented in XML format, while the right side shows the same data represented in graph format. Nodes in the graph that share a color with elements in the XML diagram represent the same object. Relationships between objects in the graph are shown as edges, representing membership or positional relationships. This transformation provides a more scalable and efficient framework for analyzing and visualizing traffic data. Further extensions can be added to the node and edge attributes in the graph format to enable more advanced transportation modeling and analysis.

\section{NUMERICAL EXPERIMENTS}
\subsection{Experiment design}

In this section, a numerical experiment was designed and conducted for the method proposed in this paper. Firstly, a simple scenario in the traffic simulation field was designed. This traffic simulation scenario includes road elements, which have an id attribute and several lanes with speed limits. It also includes vehicle elements, which have an id attribute and are classified into autonomous and human-driven vehicles travelling at different speeds on a lane of a certain road.

\begin{figure}[htbp]
\centering
\includegraphics[width=0.7\linewidth]{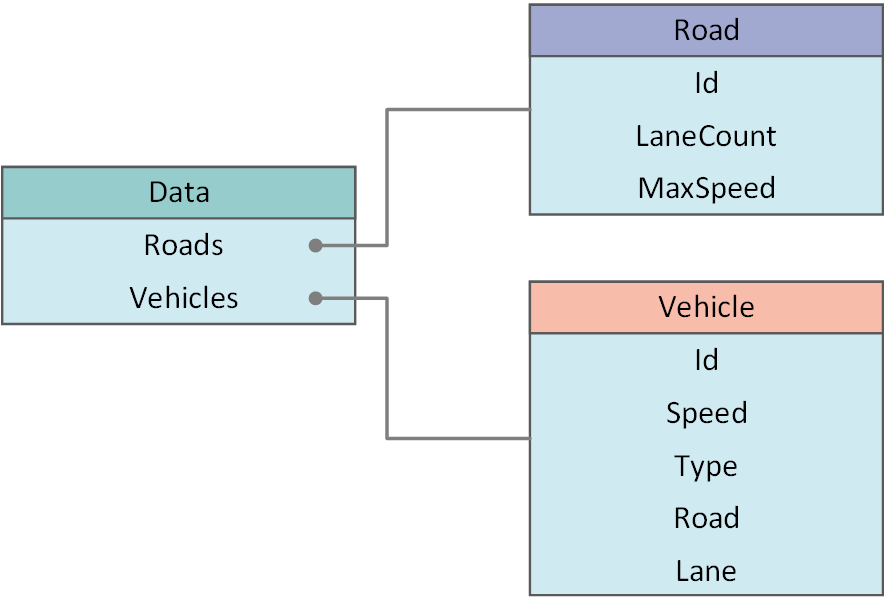}
\caption{XML design in numerical experiments}
\label{fig_xmldef}
\end{figure}

\begin{figure}[htbp]
\centering
\includegraphics[width=0.7\linewidth]{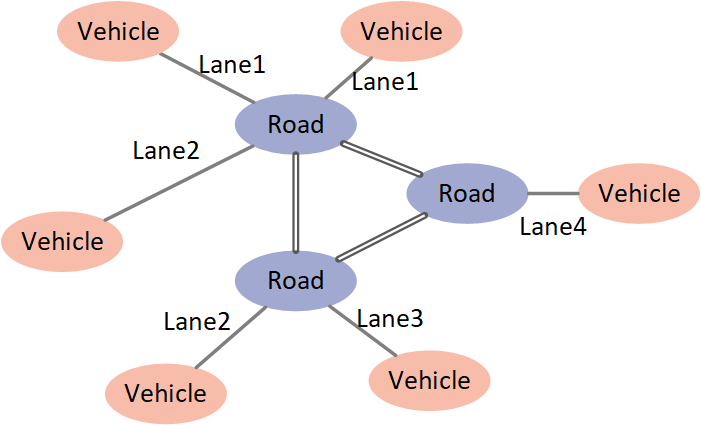}
\caption{Graph design in numerical experiments}
\label{fig_graphdef}
\end{figure}

Random scenarios were generated, comprising $m$ roads with $n$ cars on each road. The scene information was saved in XML format, and the design details of the XML are shown in Fig. \ref{fig_xmldef}. The IR convert method was used to preprocess the XML file into the Graph format, with the Graph design shown in Fig. \ref{fig_graphdef}. We performed a common traffic application task of querying the IDs of vehicles on a specific road using both XML and Graph data formats. We then conducted a comparative analysis of the respective query times for each format.

\subsection{Theoretical analysis}

In the first experiment, we directly utilized XML data to conduct vehicle ID queries. The program began with loading and parsing the given XML file, which was parsed into a two-layer structure tree without further processing, where all roads and vehicles were stored in separate lists. Afterwards, an empty list was created to store the vehicle IDs on the specified road. Subsequently, the program traversed all vehicle elements in the vehicle list. For each vehicle, the program extracted its ID, position, and type attributes, and checked whether it was located on the specified road. If so, the ID of the vehicle was added to the list. Finally, the program returned the list of vehicle IDs. There were a total of $m$ roads, each with $n$ cars, resulting in a total of $mn$ cars. Therefore, the time complexity of this algorithm is $O(mn)$.

In the second experiment, we used converted and processed graph-based IR data to conduct vehicle ID queries. The program first loaded and parsed the given graph file, which contained road and vehicle nodes, with each edge between them representing the position of the vehicle on the road, and attributes representing the lane. The nodes and edges were stored in a hash table. Next, an empty list was created to store the vehicle IDs on the specified road. The program used the hash table to look up the queried road node. Then, it traversed all adjacent nodes of the road node. For each adjacent node, the program checked whether it belonged to the vehicle type. If so, the ID of the vehicle node was added to the list. Finally, the program returned the list of vehicle IDs. As the node information is typically stored in the hash map of a Graph database, the time complexity of this step is $O(1)$. Next, the program traversed the edges to find all adjacent nodes of the queried road node and determined if they were vehicles, with a time complexity of approximately $O(n)$. Therefore, the total time complexity of this algorithm is $O(n)$.

\subsection{Experiment result}

The current experiment was conducted on a laptop with an i9-13980h CPU and 64GB of memory.

\begin{figure}[htb]
\centering
\includegraphics[width=0.8\linewidth]{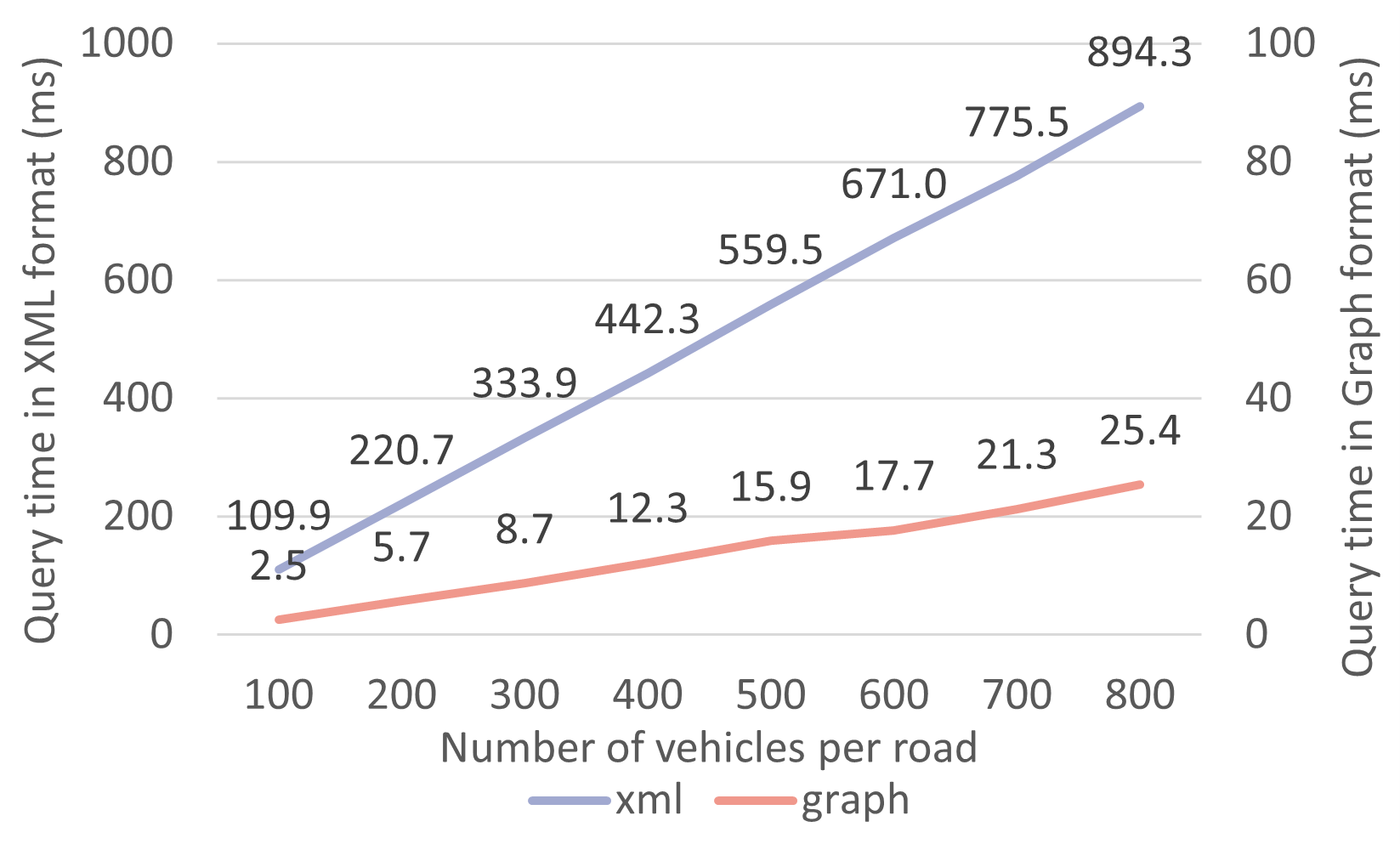}
\caption{Same road number, different vehicle density experiment}
\label{fig_vehiclenum}
\end{figure}

The first experiment involved 8 groups with varying vehicle densities but the same number of roads. Specifically, 100 random roads were generated, each containing 100, 200,..., 800 vehicles. Both XML and Graph data formats were used for querying vehicle ids on a certain road. The average query time per road was recorded and the results are presented in the graph below. It can be observed that the query time for both methods increases with the vehicle density (n). However, the Graph data format exhibited a significantly faster query time, about 40 times faster than the XML data format, as shown in Fig.\ref{fig_vehiclenum}.

\begin{figure}[!hb]
\centering
\includegraphics[width=0.8\linewidth]{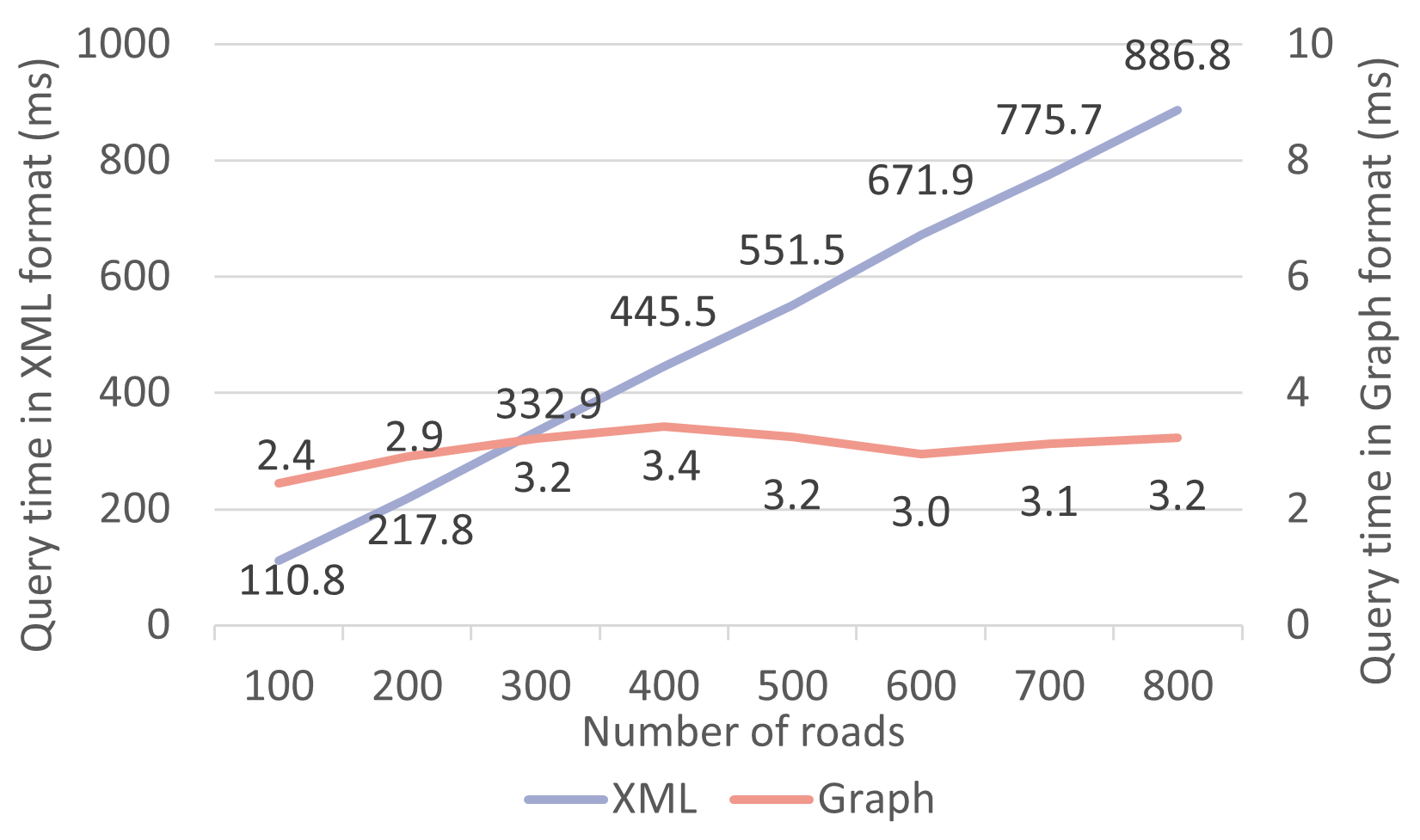}
\caption{Same vehicle density, different number of roads}
\label{fig_roadnum}
\end{figure}

The second experiment involved 8 groups with the same vehicle density but varying numbers of roads. Specifically, 100, 200, ..., 800 random roads were generated, each containing 100 vehicles. Both XML and Graph data formats were used for querying vehicle ids on a certain road. The average query time per road was recorded and the results are presented in the graph below. It can be observed that the query time for XML data format increases linearly with the number of roads, while the query time for Graph data format is independent of the number of roads. Furthermore, the superior performance of Graph data format is evident as the query time is 46 to 277 times faster than XML data format, with the advantage of using Graph data format becoming more pronounced as the number of roads increases, as shown in Fig. \ref{fig_roadnum}.

The numerical experiment results indicate that the query time for XML data format is linearly positively correlated with both the number of roads and vehicle density. In contrast, the query time for Graph data format is linearly positively correlated with vehicle density, but independent of the number of roads, which is consistent with the theoretical analysis of time complexity.

\section{CONCLUSIONS}

We propose a graph-based IR that can be used as a common data format for ASNL in the transportation field. This method efficiently improves the performance of data operations commonly used in traffic applications without sacrificing readability. Moreover, the use of a unified Graph data format as an IR facilitates the process of data format conversion between different software systems. We conducted a numerical experiment using vehicle query operations on roads, which is one of the commonly used data operations in traffic applications. We compared the query performance of directly using XML data format to using graph-based IR. The experimental results showed that the query performance using graph-based IR data format was significantly improved.

In the future, based on the proposed graph-based IR, we aim to develop a set of broadly applicable data description formats and algorithm designs that can cover the entire transportation scenario, including macro, micro, and in-vehicle perspectives, to meet the needs of handling data and interaction behavior information across multiple domains and scales, more importantly, it is adapted to the requirements of high performance ASNL. 




\section{Code availability}
The code of IR for ASNL was implemented in Python, using the graph data processing framework of Networkx. Code and scripts reproducing the experiments of this paper are available at {https://github.com/PJSAC/IR4ASNL}.





\bibliographystyle{IEEEtran}
\bibliography{./IEEEabrv, ./IEEEexample}
\end{CJK}
\end{document}